\documentclass[prb,showpacs,preprintnumbers,amsmath,amssymb,twocolumn,superscriptaddress,longbibliography]{revtex4-1}

\usepackage[dvipdfmx]{graphicx}
\usepackage{dcolumn}
\usepackage{bm}
\usepackage{color}

\usepackage{amsmath}	
\usepackage{amsfonts}
\usepackage{braket}
\usepackage{amsthm}

\usepackage{bbm}

\begin{document}

\title{
Spin susceptibility in quasicrystal superconductor with spin-orbit interactions
}

\author{Yasuhiro Tada}
\email[]{ytada@hiroshima-u.ac.jp}
\affiliation{
Quantum Matter Program, Graduate School of Advanced Science and Engineering, Hiroshima University,
Higashihiroshima, Hiroshima 739-8530, Japan}
\affiliation{Institute for Solid State Physics, University of Tokyo, Kashiwa 277-8581, Japan}


\begin{abstract}
We study impacts of spin-orbit interactions on the spin susceptibility in quasicrystal superconductors,
motivated by the anomolous superconducting properties in the van der Waals quasicrystal Ta$_{1.6}$Te under magnetic fields.
We consider the Penrose tiling model with $s$-wave pairing as a representative system 
and include anisotropic spin-orbit interactions allowed for the 
quasicrystal structure.
It is shown that the spin-momentum locking locally takes place in the quasicrystal, and electron motion and spin directions
are tightly connected at each spatial position in the real space.
As a result, the spin susceptibility is enhanced for both in-plane and out-of-plane magnetic fields 
in the presence of a Rashba-type spin-orbit interaction.
For an Ising-type spin-orbit interaction, the spin susceptibility only for in-plane magnetic fields is  increased,
while the out-of-plane spin susceptibility is almost unchanged.
When there are both kinds of the spin-orbit interactions, temperature dependence of the spin susceptibility for any field directions 
is suppressed.
\end{abstract}

\maketitle

\section{Introduction}
\label{sec:introduction}
Superconductivity is a coherently Cooper paired state and is realized in various crystalline systems.
However, the phase coherence can survive even without clean crystal periodicity and indeed is possible
in a dirty system~\cite{Anderson1959}.
Furthermore, superconductivity can be realized in a quasicrystal which possesses rotaion symmetry but does not
have any translation symmetry.
This was first established experimentally in Al-Zn-Mg based quasicrystal, 
where the superconducting transition temperature is $T_c\simeq 0.05$K~\cite{Kamiya2018}.
Superconductivity with a high transition temperature $T_c\simeq1$K was found in the van der Waals layered
quasicrystal Ta$_{1.6}$Te~\cite{Conrad1998,Tokumoto2024}.
In this system, the upper critical field exhibits linear temperature dependence down to 0.04K 
and is anomalously large, $H_{c2}(0)\simeq 4$T,
and it well exceeds the Pauli limit for spin singlet superconductivity, $H_{c2}^{\rm Pauli}\simeq 1.8$T~\cite{Terashima2024}.
Furthermore, the Knight shift in the NMR experiment shows very weak temperature dependence down to $\sim 0.1$K,
which means that spin magnetism of Ta$_{1.6}$Te is highly non-trivial~\cite{Matsudaira2025}.
On the other hand, the observed $T_1$ exhibits a coherence peak around $T\simeq T_c$ and exponential temperature dependence
at low temperature, 
which would be a clear signature for fully gapped $s$-wave
superconductivity.
In addition, the observed transition temerature $T_c\simeq 1$K in the Cu-doped $({\rm Ta}_{0.95}{\rm Cu}_{0.05})_{1.6}$Te 
is unchanged from that in the undoped ${\rm Ta}_{1.6}$Te, suggesting $s$-wave pairing which is robust to impurities.
Although there have been fundamental theoretical studies for quasicrystal superconductors~\cite{Sakai2017,Sakai2019,Takemori2020,Cao2020,Nagai2022},
the anomalous experimental observations cannot be fully understood within the existing theories.

It is known that, in general periodic crystal systems, 
superconductivity is robust to magnetic fields when spins and momenta of electrons on the Fermi surface are locked 
each other due to anisotropic
spin-orbit interactions~\cite{Sigrist2012}.
The spin-momentum locking leads to spin textures formed in the $k$-space. 
Cooper pairing occurs for such spin-locked electrons and therefore it is robust to Zeeman magnetic fields~\cite{Frigeri2004,Frigeri2004NJP}.
For example, the upper critical field parallel to the $c$-axis 
well exceeds the Pauli limit in Rashba superconductors such as CePt$_3$Si~\cite{Bauer2004,Yasuda2004}, CeRhSi$_3$~\cite{Kimura2005,Kimura2007},
and CeIrSi$_3$~\cite{Sugitani2006,Settai2008}.
$H_{c2}^{\perp z}$s perpendicular to the $c$-axis are also larger than $H_{c2}^{\rm Pauli}$ and theoretical studies taking 
spin-orbit interactions into account well explain such behaviors~\cite{Tada2008,Tada2010}.
In the few-layer van der Waals materials, NbSe$_2$ and MoS$_2$, 
the in-plane upper critical fields reach $H_{c2}^{\perp z}\sim 7H_{c2}^{\rm Pauli}$
and these colossal enhancements are attributed to the Ising-type anisotropic spin-orbit 
interactions~\cite{Lu2015,Xiaoxiang2016,Saito2016,Wang2021}.
On the other hand,
the $H_{c2}^{\parallel z}$ parallel to the $c$-axis is suppressed by the orbital deparing effect and is smaller than $H_{c2}^{\rm Pauli}$.
Similar anomalous behaviors of the upper critical fields have been observed in the SrTiO$_3$/LaAlO$_3$ interface~\cite{Shalom2010} and 
the SrTiO$_3$ surface~\cite{Ueno2014}, which was also understood based on the spin-orbit interaction~\cite{Nakamura2013}.
Furthermore, an exchange field in a ferromagnetically ordered system can also play a smilar role to those of 
anisotropic spin-orbit interactions, which is important in ferromagnetic superconductors~\cite{Mineev2010,Hattori2012,Tada2013,Tada2017}.

In the quasicrystal superconductor Ta$_{1.6}$Te, there would be no global inversion symmetry because its approximants lack 
mirror symmetry along the $c$-axis as pointed out in the experimental works~\cite{Matsudaira2025,Conrad2002,Cain2020}. 
This could lead to a Rashba-type spin-orbit interaction.
In addition, Ta$_{1.6}$Te contains many bonds around which there is no inversion symmetry and
anisotropic spin-orbit interactions can arise also from such locally non-centrosymmetric structures~\cite{Maruyama2012}.
These anisoctropic spin-orbit interactions are expected to enhance the spin susceptibility and consequently 
stability of superconductivity in a quasicrystal 
under magnetic fields as in
periodic crystalline superconductors.
However,
it is not trivial whether or not the spin-momentum locking mechanism indeed works in a quasicrystal superconductor, 
since there is no translation symmetry
and the Cooper pairing of $(\vec{k},-\vec{k})$ momenta is not a good description.
The effectiveness of the spin-momentum locking in Ta$_{1.6}$Te was implicitly assumed in the discussions on the experiments
as a possible origin of the anomalous behaviors,
but its validity should be theoretically examined to develop a firm understanding.

In this study, motivated by the experimental findings of the weak temperature dependence of the Knight shift and 
the robust superconductivity in Ta$_{1.6}$Te under magnetic fields,
we investigate the spin susceptibility and consider 
effects of the anisotropic spin-orbit interactions due to both global and local inversion asymmetries in a quasicrystal superconductor.
We consider the Penrose tiling model with $s$-wave pairing as a simple representative model for two-dimensional quasicrystal systems.
It is shown that spin-momentum locking takes place locally for each spatial position even in the absence of the translation symmetry.
As a result, 
the spin susceptibility $\chi$ for both in-plane (e.g. parallel to the $x$-axis) and out-of-plane ($z$-axis) magnetic fields 
is strongly enhanced by the Rashba spin-orbit interaction.
Futhermore, $\chi^{xx}$ for a $x$-axis magnetic field shows only weak temperature dependence 
when there is a sufficiently strong  Ising spin-orbit interaction.

\section{Model}
\label{sec:model}
\subsection{Hamiltonian and spin-orbit interactions}
\label{sec:hamiltonian}
\begin{figure}[htb]
\includegraphics[width=7.0cm]{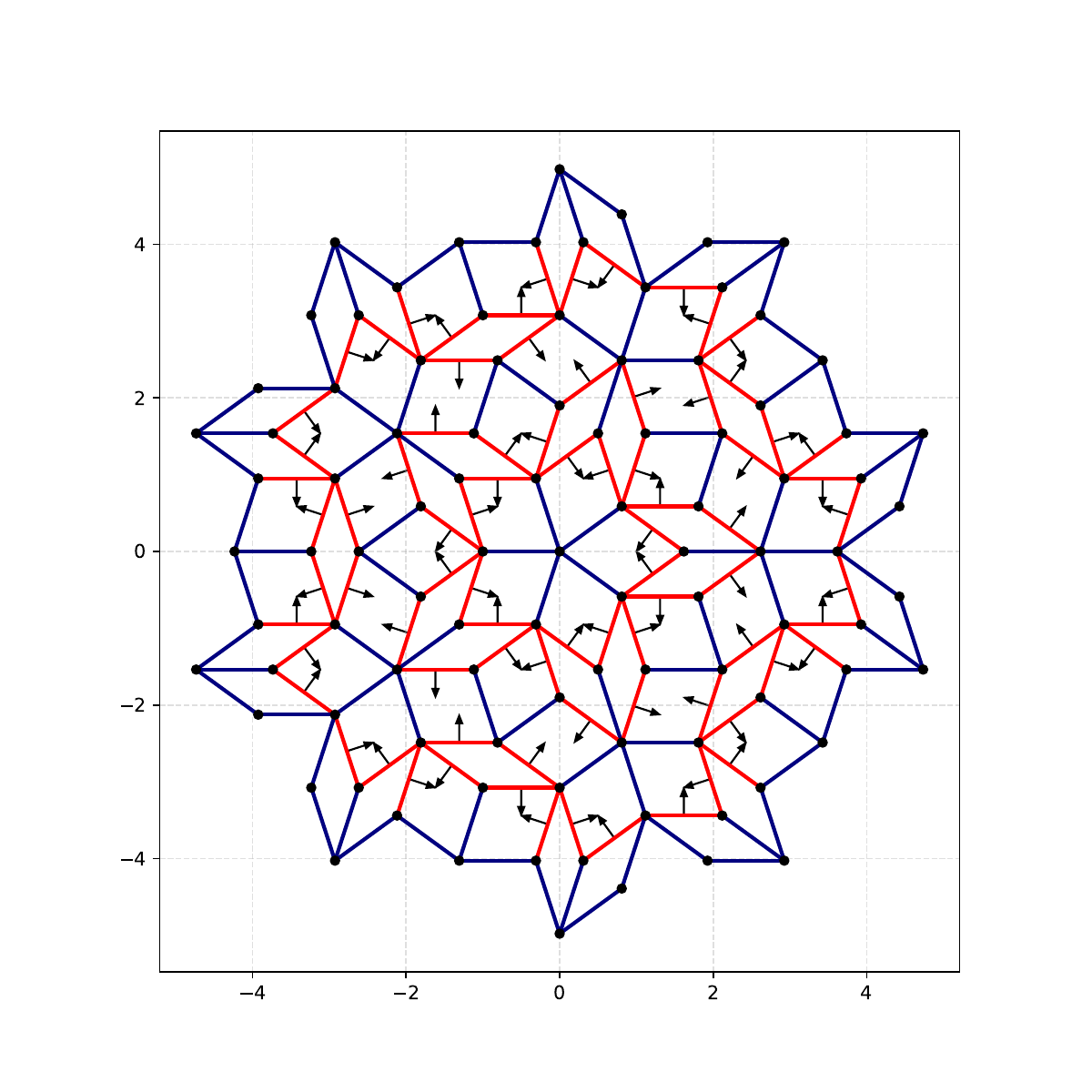}
\caption{The edges of the rhombuses and the normal vector $\vec{n}_{ij}$ for the bonds (red) shared by both large and small rhombuses.
$\vec{n}_{ij}=0$ for the bonds (navy) which are not shared by large and small rhombuses. 
The system size $N=96$ in this figure is chosen for clear visibility.
For a large system, the ratio (the number of the red bonds)/(the total number of all the bonds) is $\simeq 0.6$.
The $(x,y)$ coordinate is also shown.
}
\label{fig:nij}
\end{figure}

We consider $s$-wave superconductivity on the Penrose tiling as a simple representative model for two-dimensional quasicrystals.
The Hamiltonian is 
\begin{align}
H&=-t\sum_{ij,s}c_{is}^{\dagger}c_{js} + \sum_{i,s}\frac{\Delta_{ss'}}{2}\left(c^{\dagger}_{is}c^{\dagger}_{is'}+({\rm h.c.})\right) \nonumber\\
&\quad + i\alpha_{\perp}\sum_{ij,ss'}\vec{g}^{\perp}_{ij}\cdot \vec{\sigma}_{ss'}c_{is}^{\dagger}c_{js'}
+i\alpha_{\parallel} \sum_{ij,ss'}\vec{g}^{\parallel}_{ij}\cdot \vec{\sigma}_{ss'}c_{is}^{\dagger}c_{js'},
\label{eq:hamiltonian}
\end{align}
where we consider hopping and spin-orbit interactions only for the edges of the rhombuses (Fig.~\ref{fig:nij}). 
$t=1$ is the energy unit.
The spin index is $s=\uparrow,\downarrow$
and $\vec{\sigma}$ is the Pauli matrix.
The average particle filling $n_e=1/N\sum_{is}\braket{c^{\dagger}_{is}c_{is}}$ is determined by the chemical potential, 
where $N$ is the total number of sites.
To focus on impacts of the spin-orbit interactions,
we have assumed that the superconducting gap function is uniform over the whole system, although 
it is non-uniform in microscopic calculations~\cite{Sakai2017,Sakai2019,Takemori2020,Cao2020,Nagai2022}.
The gap function describes the on-site $s$-wave pairing, $\Delta_{\uparrow\downarrow}=-\Delta_{\downarrow\uparrow}=\Delta$ and 
$\Delta_{\uparrow\uparrow}=\Delta_{\downarrow\downarrow}=0$.
The $s$-wave gap symmetry is consistent with the NMR experiments in $({\rm Ta}_{0.95}{\rm Cu}_{0.05})_{1.6}$Te 
where the relaxation rate $T_1$ exihibits
a coherence peak around $T\simeq T_c$ and an exponential temperature dependence at low temperature 
in addition to the robust $T_c$ under the Cu doping~\cite{Matsudaira2025}.

For the spin-orbit interactions, we follow the simple phenomenological argument in translationally symmetric systems that 
they are given by $\vec{g}\cdot\vec{\sigma}$ with $\vec{g}=\vec{v}\times\vec{E}$
where $\vec{v}$ is an electron velocity and $\vec{E}$ is an electric field.
Such a simple argument is known to be qualitatively consistent with detailed analyses of multi-orbital systems
in some parameter regimes
where anisotropic spin-orbit interactions arise from interplay between the on-site spin-orbit interaction and 
the orbital hybridization~\cite{Nakamura2013,Yanase2013}.
Spin-orbit interactions in the Penrose tiling can be discussed in a similar manner within the phenomenological approach.
In our system, the vectors $\vec{g}^{\perp}_{ij}$ and $\vec{g}^{\parallel}_{ij}$ describe the Rashba and Ising spin-orbit interactions, respectively.
They are given by
\begin{align}
\vec{g}^{\perp}_{ij}&=\vec{r}_{ij}\times\hat{z}, \\
\vec{g}^{\parallel}_{ij}&=\vec{r}_{ij}\times\vec{n}_{ij},
\end{align}
where $\vec{r}_{ij}$ is the vector connecting the two sites $i,j$.
The $g$-vector lies within the $xy$-plane for the Rashba interaction, while it is parallel to the $z$-axis for the Ising interaction.
The unit vector $\vec{n}_{ij}$ is defined as follows.
In the Penrose tiling, there are two kinds of the rhombuses, namely, small ones and large ones.
For a nearest neighbor bond which is shared by both kinds of rhombuses, there will be a local effective electric field perpendicular to
the bond due to the lack of the local inversion symmetry. 
We denote the direction of the electric field by $\vec{n}_{ij}$ for the bonds connecting the two sites $i,j$,
and assume for simplicity that it points from the small rhombus to the large rhombus.
A configuration of $\vec{n}_{ij}$ is shown in Fig.~\ref{fig:nij}. 
Note that the Ising spin-orbit interaction works only for a limited number of bonds and the ratio
(the number of the bonds with $\vec{g}_{ij}^{\parallel}\neq0$)/(the total number of all the bonds) is numerically
$\simeq 0.6$ for a large system size $N=4221$ which we use in Sec.~\ref{sec:spin_susceptibility} for calculations of the spin susceptibility.

We expect that a single layer of Ta$_{1.6}$Te contains several kinds of clusters of Ta and Te,
and the layer exhibits a ripple-like structure, since the local structure would be similar to its approximants~\cite{Conrad2002,Cain2020}.
Therefore, the effective local electric field $\vec{E}$ would be non-uniform and may point to different directions
depending on spatial positions.
Especially, the Rashba $g$-vector $\vec{g}^{\perp}_{ij}$ could be flipped at different spatial positions.
Although realistic position dependence of the $g$-vectors would be important in understanding Ta$_{1.6}$Te,
we assume simple $g$-vectors to understand basic properties of a spin-orbit coupled quasicrystal superconductor.

\subsection{Local spin-momentum locking}
\label{sec:locking}
As mentioned in the introduction, the spin-momentum locking is a key for the enhancement of the spin susceptibility
in periodic crystalline superconductors with spin-orbit interactions~\cite{Sigrist2012,Frigeri2004,Frigeri2004NJP}.
Although momentum in the $k$-space is not well defined on a quasicrystal,
the electron motion at each spatial position has a clear physical meaning when a particle is regarded as a point-like
object in the real space.
For example, the electron motion on a quasicrystal can be characterized locally by the correlation function $\braket{c^{\dagger}_{is}c_{js}}$.
In the present system, electron motion and spin directions are tightly connected due to the spin-orbit interaction.
To see this, we focus on the non-superconducting state ($\Delta=0$) and define inter-site spin correlations
\begin{align}
\vec{s}_{ij}=\sum_{ss'}i\vec{\sigma}_{ss'}\braket{c^{\dagger}_{is}c_{js'}-c^{\dagger}_{js}c_{is'}}
\end{align}
for each nearest neighbor bond.
This quantity is real and time-reversal even, and is vanishing in the absence of the spin-orbit interactions.

\begin{figure}[htb]
\includegraphics[width=7.0cm]{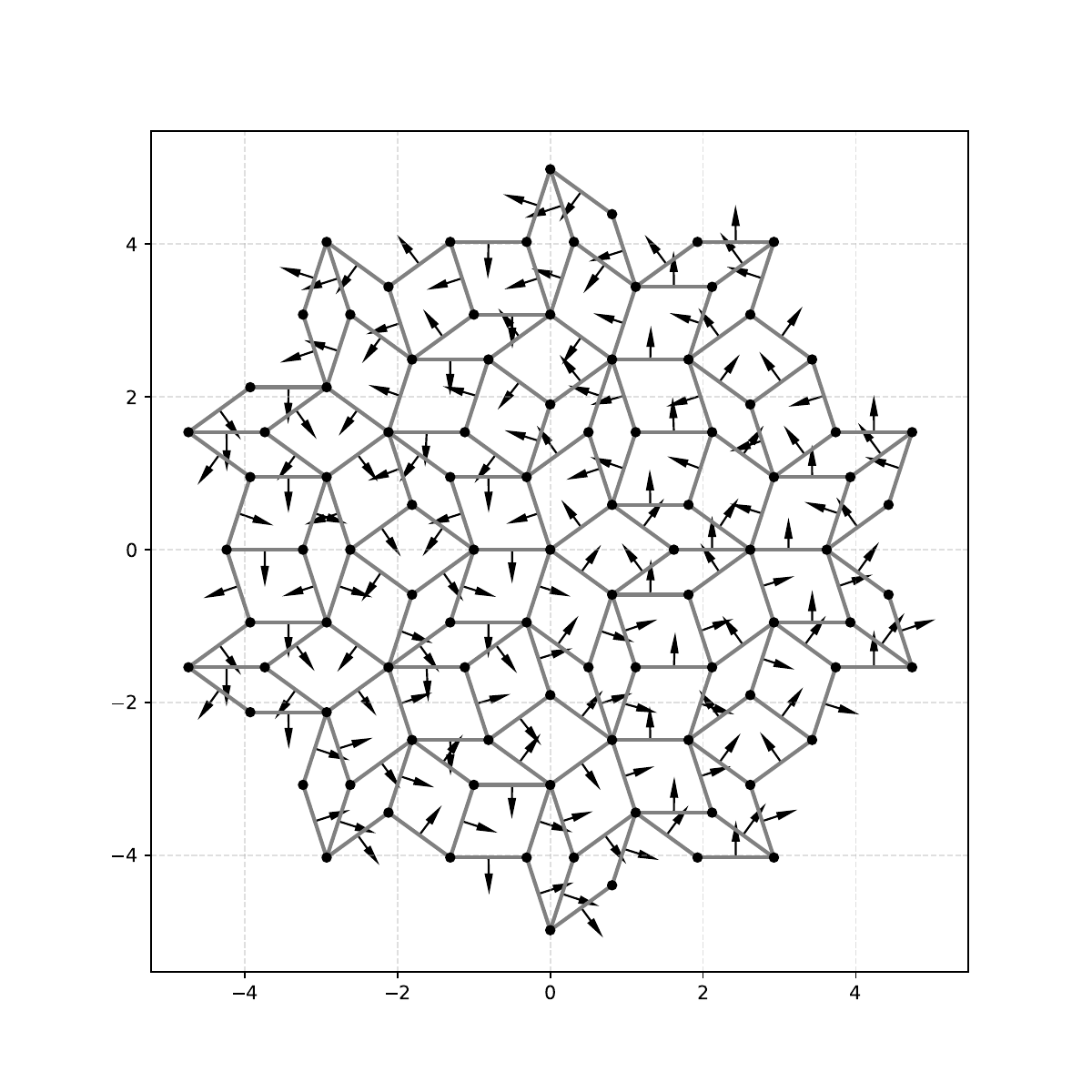}
\caption{The inter-site spin correlation 
$\vec{s}_{ij}^{\perp}=(s_{ij}^x,s_{ij}^y)$ in the Rashba spin-orbit coupled system ($\alpha_{\perp}=0.2, \alpha_{\parallel}=0$),
characterizing the local spin-momentum locking.
}
\label{fig:S_Rashba}
\end{figure}

\begin{figure}[htb]
\includegraphics[width=7.0cm]{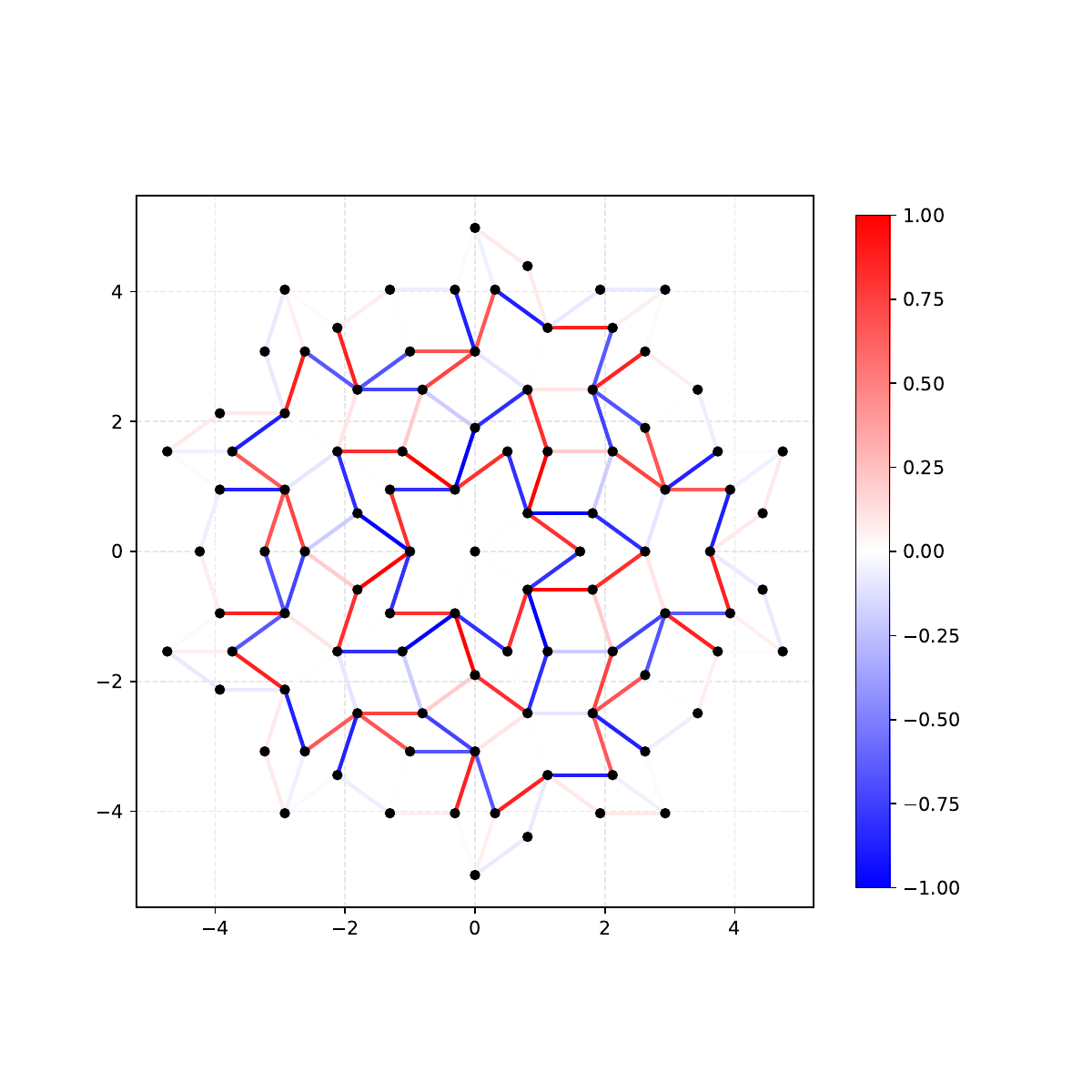}
\caption{The inter-site spin correlation
$s_{ij}^z$ (normalized by ${\rm max}_{ij}|\vec{s}_{ij}|$) in the Ising spin-orbit coupled system ($\alpha_{\perp}=0, \alpha_{\parallel}=0.2$),
characterizing the local spin-momentum locking.
}
\label{fig:S_Ising}
\end{figure}

We show typical behaviors of $\vec{s}_{ij}$ in Figs.~\ref{fig:S_Rashba} and \ref{fig:S_Ising}, where the system size $N=96$ is chosen
for clear visibility.
For the system only with the Rashba spin-orbit interaction ($\alpha_{\perp}\neq0,\alpha_{\parallel}=0$), the spin correlation $\vec{s}_{ij}$
lies almost within the $xy$-plane and it is nonzero everywhere.
On the other hand, for the system only with the Ising spin-orbit interaction ($\alpha_{\perp}=0,\alpha_{\parallel}\neq0$),
$\vec{s}_{ij}$ is parallel to the $z$-axis and it has a large magnitude only for the bonds where $\vec{g}^{\parallel}_{ij}\neq0$.
In any case, the spin-momentum locking takes place locally at each bond in the real space.
This is regarded as a generalization of the spin-momentum locking in a translationally symmetric system,
where the spin-momentum locking is related to a spin texture on the Fermi surface.

When a small Zeeman magnetic field $\vec{h}$ is applied, spins on the bond $(i,j)$ are fixed as $\vec{s}_{ij}$ and cannot point to the direction
of the magnetic field if $\vec{h}$ is not parallel to $\vec{s}_{ij}$.
In the Rashba spin-orbit coupled system, this condition is satisfied for most of the bonds 
for an in-plane magnetic field (e.g. $\vec{h}\parallel \hat{x}$), because of the complicated lattice structure characteristic of a quasicrystal.
For an out-of-plane  magnetic field, $\vec{h}$ is not prallel to $\vec{s}_{ij}$ for any bonds.
Therefore, the Rashba superconductivity is expected to be robust to magnetic fields in any directions.
On the other hand, in the Ising system, $\vec{s}_{ij}$ on all the bonds are not parallel to an in-plane magnetic field,
while they are all parallel to an out-of-plane magnetic field.
This implies that the Ising superconductivity will be robust to an in-plane magnetic field and is unstable to an out-of-plane magnetic field.
Therefore,
it is expected that, thanks to the local spin-momentum locking, 
the spin susceptibility can be strongly enhanced in the superconducting state
if the spin-orbit interaction is sufficiently larger than the gap amplitude $\Delta$.
This would be relevant for understanding the anomalous observations in Ta$_{1.6}$Te as pointed out in
the experimental works~\cite{Terashima2024,Matsudaira2025}.
We will discuss the spin susceptibility in the next section.


\section{Spin susceptibility}
\label{sec:spin_susceptibility}
We discuss the spin susceptibility in this section.
To evaluate the spin susceptibility, we add the Zeeman term
\begin{align}
H_{\rm Zeeman}= -\vec{h}\cdot \sum_{iss'}\vec{\sigma}_{ss'}c^{\dagger}_{is}c_{is'},
\end{align}
where we have simply assumed that the $g$-factor is 2.
Then, the $\mu$ component of the spin susceptibility ($\mu=x,z$) is given by
\begin{align}
\chi^{\mu\mu}=\lim_{h\to0}\frac{m_{\mu}}{h_{\mu}},
\label{eq:chi}
\end{align}
where $m_{\mu}=1/N\sum_{i,ss'}\sigma^{\mu}_{ss'}\braket{c^{\dagger}_{is}c_{is'}}$ is the magnetization averaged over the system.
The $(x,y)$ coordinate is defined in Fig.~\ref{fig:nij}.
By diagonalizing the total Hamiltonian under the magnetic field with use of the Bogoliubov transformation,
we can take the thermal average of the operators to evaluate $m_{\mu}$.
It is noted that the spin magnetization depends on spatial positions and correspondingly the local spin susceptibility
at each site is also position dependent.
We expect that the Knight shift in Ta$_{1.6}$Te in the polycrystal sample~\cite{Matsudaira2025}
would be a spatially averaged result corresponding to the $\chi^{\mu\mu}$ in Eq.~\eqref{eq:chi}.
 
In the present study, 
we assume the standard weak coupling temperature dependence for the gap function, 
$\Delta=1.76T_c\tanh(1.74\sqrt{T_c/T-1})$,
where $T_c$ is the transition temperature.
The superconducting transition temperature is a parameter of the model and is
set as $T_c=0.1$. Such a value of $T_c$ has been obtained 
in the attractive Hubbard model on the Penrose tiling without spin-orbit interactions~\cite{Sakai2017,Sakai2019}.
We denote the spin susceptibility as $\chi^{zz}_s=\chi^{zz}(T)$ in the superconducting state
and $\chi^{zz}_n=\chi^{zz}(T_c)$ in the normal state.
We discuss effects of the Rashba and Ising spin-orbit interactions separately, and the system with both of these interactions
are briefly discussed in Appendix~\ref{sec:rashbaising},
where effects of the two spin-orbit interactions simply add up.
The average filling is fixed as $n_e=1/N\sum_{is}\braket{c^{\dagger}_{is}c_{is}}=0.5$.
We perform calculations up to the system size $N=4221$, and it turns out that finite size effects are negligibly small.
In the following, the system size is fixed as $N=4221$.

\subsection{Rashba spin-orbit interaction}
\label{sec:rashba}
\begin{figure}[htb]
\includegraphics[width=6.0cm]{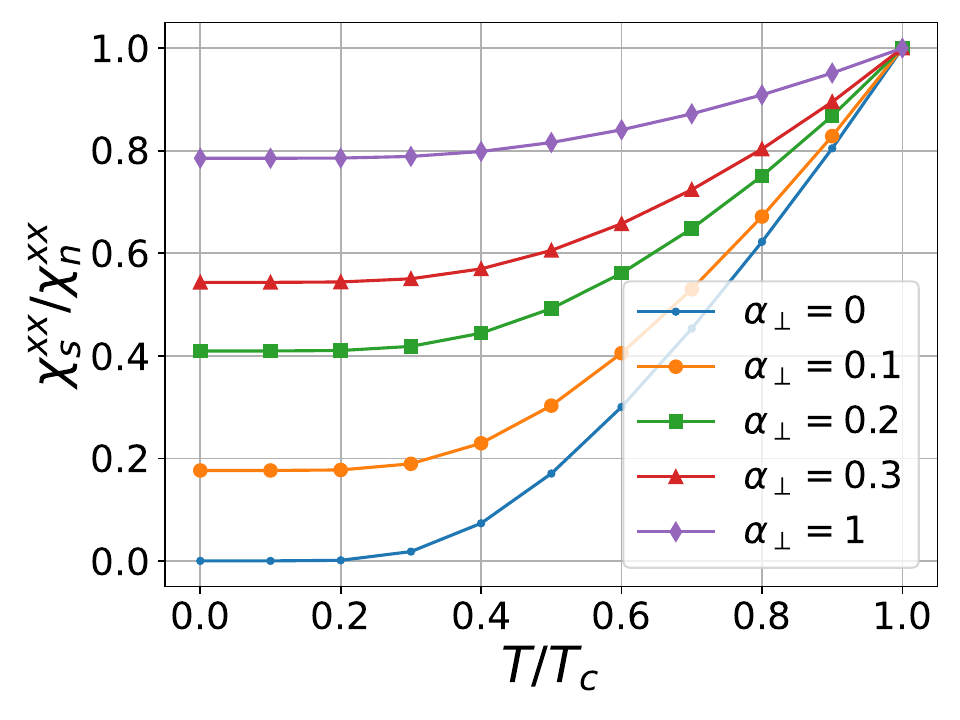}
\caption{Spin susceptibility $\chi^{xx}$ for several values of the Rashba spin-orbit coupling $\alpha_{\perp}$
The Ising spin-orbit coupling is $\alpha_{\parallel}=0$ and
the transition temperature is $T_c=0.1$.
}
\label{fig:chix_aR}
\end{figure}

\begin{figure}[htb]
\includegraphics[width=6.0cm]{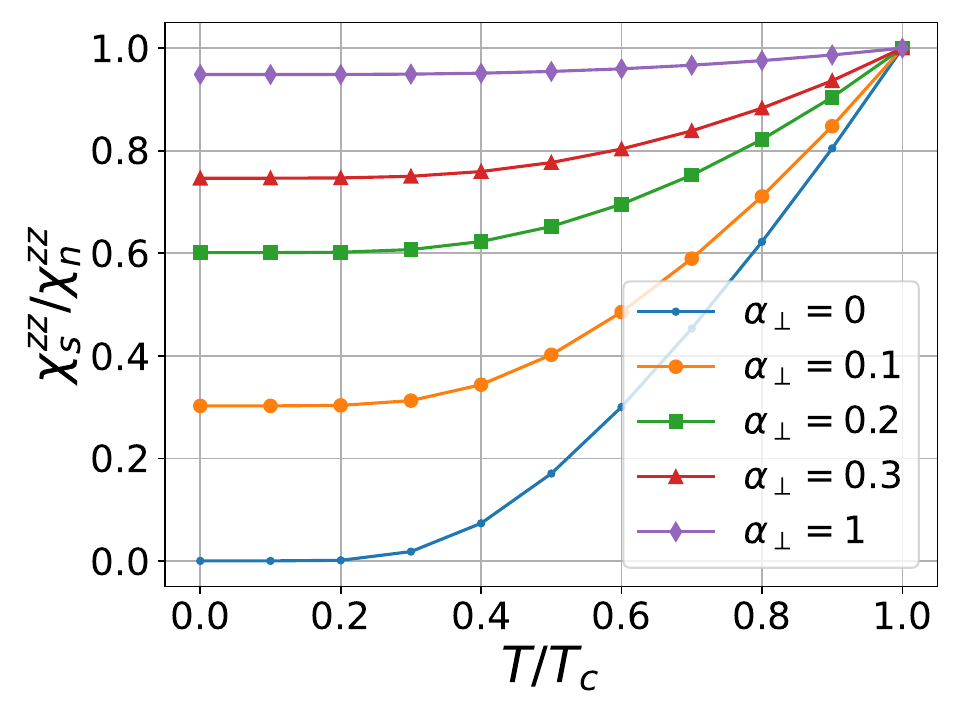}
\caption{Spin susceptibility $\chi^{zz}$ for several values of the Rashba spin-orbit coupling $\alpha_{\perp}$. 
The Ising spin-orbit coupling is $\alpha_{\parallel}=0$ and
the transition temperature is $T_c=0.1$.
}
\label{fig:chiz_aR}
\end{figure}

We first consider effects of the Rashba spin-orbit interaction with the coupling constant $\alpha_{\perp}$ 
and set $\alpha_{\parallel}=0$ for the Ising interaction.
As seen in Figs.~\ref{fig:chix_aR} and \ref{fig:chiz_aR}, the spin susceptibility is enhanced by the Rashba interaction
for both directions (parallel to the $x$- and $z$-axes) of the magnetic fields.
This means that the local spin-momentum locking mechanism indeed works even in the absence of the translation symmetry.
An applied small magnetic field is irrelevant for the superconductivity because the spins are tied to the electron motion.

Interestingly, $\chi^{xx}$ is strongly enhanced by the Rashba spin-orbit interaction. 
This is because the spin-momentum locking works for a large number of bonds due to the complicated lattice structure
as mentioned in the previous section.
The strong enhancement can be compared with $\chi^{xx}$ in the canonical Rashba superconductor where $\chi^{xx}=\chi^{zz}/2$ at $T=0$~\cite{Frigeri2004NJP}.
Although a large $\chi^{xx}$ and robust superconductivity for in-plane mangetic fields
can be seen also in some periodic systems such as 
a strongly correlated system~\cite{Fujimoto2007} and a multi-orbital system~\cite{Nakamura2013},
the large $\chi^{xx}$ in the present system purely arises from the quasicrystal structure.
When $\alpha_{\perp}\gg \Delta$,
the spin susceptibility $\chi^{zz}$ for the out-of-plane magnetic field only slightly decreases when temperature is reduced.
Therefore, it is considered that, if the Rashba-type spin-orbit interaction is present and is sufficiently large compared to the transition temperature
$1$K in Ta$_{1.6}$Te,
the Knight shift will show weak temperature dependence in the superconducting state.
This will essentially apply to any superconducting gap function other than the $s$-wave pairing, although the pairing symmetry
has not been fully identified in Ta$_{1.6}$Te.
However, at the same time, it is known that some gap functions cannot be consistent with the spin-orbit interactions and 
the corresponding superconductivity is suppressed 
in periodic systems~\cite{Frigeri2004,Ramires2018}.
Similar suppression of superconductivity for some gap functions may occur also in quasicrystal superconductors
with spin-orbit interactions.


\subsection{Ising spin-orbit interaction}
\label{sec:ising}
\begin{figure}[htb]
\includegraphics[width=6.0cm]{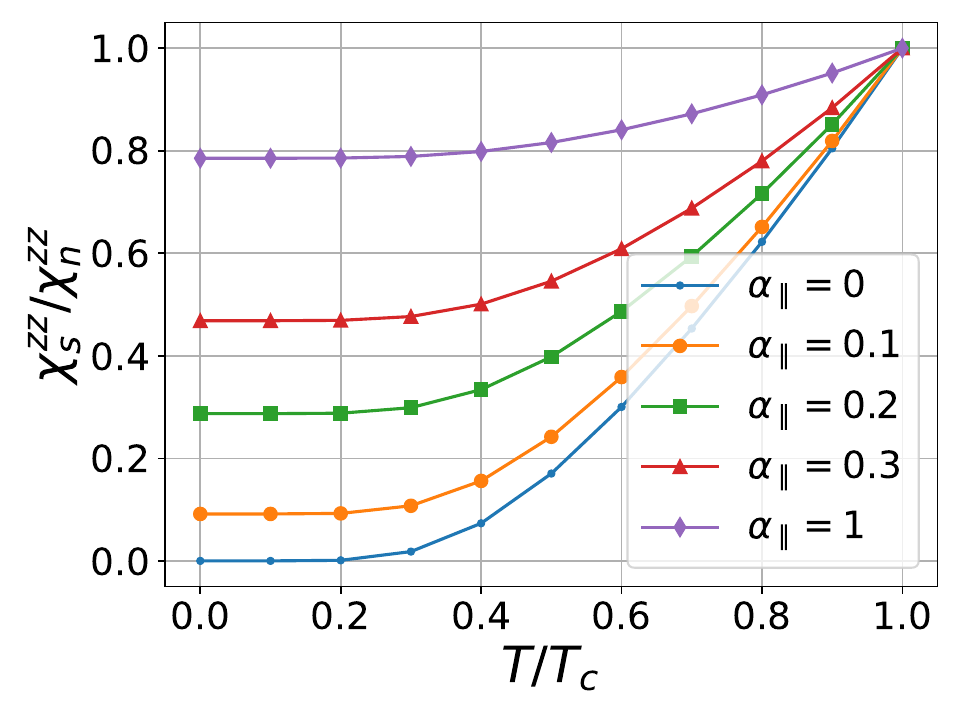}
\caption{Spin susceptibility $\chi^{xx}$ for several values of the Ising spin-orbit coupling $\alpha_{\parallel}$.
The Rashba spin-orbit coupling is $\alpha_{\perp}=0$ and
the transition temperature is $T_c=0.1$.
}
\label{fig:chix_lam}
\end{figure}

We next discuss effects of the Ising spin-orbit interaction with the coupling constant $\alpha_{\parallel}$ 
and set $\alpha_{\perp}=0$ for the Rashba interaction.
As shown in Fig.~\ref{fig:chix_lam}, the in-plane spin susceptibility $\chi^{xx}$ is enhanced by the Ising spin-orbit interaction.
The magnitude of $\chi^{xx}$ is comparable with $\chi^{zz}$ in the system with the Rashba interaction discussed in the previous section,
although there are only a restricted number ($\simeq$ 60\%) of the bonds where $\vec{g}^{\parallel}_{ij}\neq0$.
(One should compare $\chi^{xx}$ in the Ising system with $\chi^{zz}$ in the Rashba system because the same condition 
$\vec{g}_{ij}\perp\vec{h}$ uniformly
holds in both cases.)
When the spin-orbit interaction is not so large, $\alpha_{\parallel}\leq 0.2$,
the ratio $(\chi^{xx}_s/\chi^{xx}_n)_{\rm Ising}/(\chi^{zz}_s/\chi^{zz}_n)_{\rm Rashba}$ is roughly $\lesssim0.5$ corresponding to the
ratio between the spin-orbit coupled bonds and all the bonds.
Interestingly, when the Ising spin-orbit interaction
is sufficiently strong, $\alpha_{\parallel}\gg \Delta$, the ratio of the spin susceptibility approaches unity exceeding 
the naively expected value 0.6.
This means that the Ising spin-orbit interaction only on the bonds with $\vec{g}^{\parallel}_{ij}\neq0$ can determine the spin magnetic properties
of the entire system.

On the other hand, the out-of-plane spin susceptibility $\chi^{zz}$ is almost unchaged by the Ising spin-orbit interaction
over the whole temperature range $0\leq T\leq T_c$.
This is because the local spin-momentum locking (Fig.~\ref{fig:S_Ising}) is parallel to the $z$-axis and 
thus the system can be easily affected by a $z$-axis magnetic field.
The anisotropic behaviors of $\chi^{\mu\mu}$ in the present system are qualitatively similar to an Ising spin-orbit coupled system
with the translation symmetry.
Our result implies that it is difficult to understand the Knight shift in Ta$_{1.6}$Te based only on the Ising spin-orbit interaction
and there would be strong Rashba spin-orbit interactions in this material.
We note that the NMR experiment~\cite{Matsudaira2025} has been done for a polycrystal and 
the magnetic field orientation with respect to
the layered structure has not been identified.
If the crystal domain orientations are nearly random,
we expect that an averaged spin susceptibility $\chi^{\rm av}=(\chi^{xx}+\chi^{yy}+\chi^{zz})/3$ would be observed in an experiment.
In this case, $\chi^{\rm av}$ is at most $\chi^{\rm av}\simeq 2/3\chi^{xx}$ if only the Ising spin-orbit interaction exists, 
which is not well consistent with 
the weak temperature denpendence of the Knight shift observed in the experiment~\cite{Matsudaira2025}.

\begin{figure}[htb]
\includegraphics[width=6.0cm]{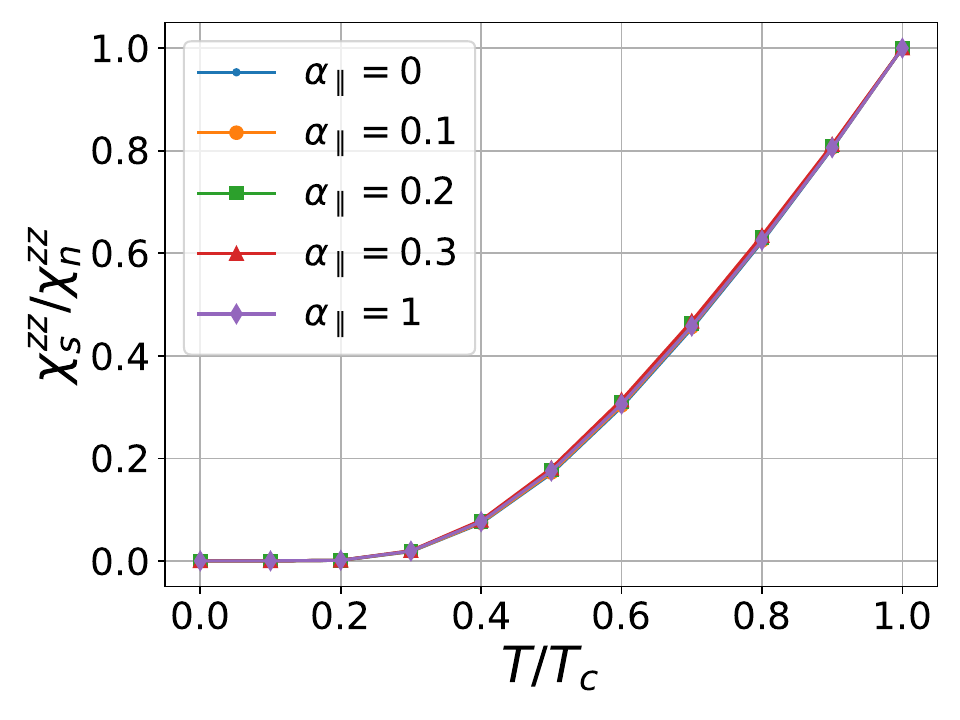}
\caption{Spin susceptibility $\chi^{zz}$ for several values of the Ising spin-orbit coupling $\alpha_{\parallel}$. 
The Rashba spin-orbit coupling is $\alpha_{\perp}=0$ and
the transition temperature is $T_c=0.1$.
}
\label{fig:chiz_lam}
\end{figure}

\section{Summary and Discussion}
\label{sec:summary}
We have discussed the spin-momentum locking and the spin susceptibility 
in the quasicrystal superconductor with the spin-orbit interaction on the Penrose tiling.
It was shown that the spin-mometum locking takes place locally at each spatial position, although there is no translation symmetry.
The inter-site spin correlation in the non-superconducting state can characterize the spin-momentum locking in the real space
and it lies almost within the $xy$-plane for the Rashba spin-orbit interaction,
while it is parallel to the $z$-axis for the Ising interaction.
As a result, the spin susceptibility is enhanced in the superconducting state.
For the Rashba spin-orbit interaction, both of $\chi^{xx}$ and $\chi^{zz}$ are increased and their magnitudes are comparable.
The relatively large ratio $\chi^{xx}/\chi^{zz}$ arises purely from the quasicrystal structure, which is distinct from
the canonical Rashba superconductors with the translational symmetry.
For the Ising interaction, $\chi^{xx}$ is strongly enhanced despite the reduced number of bonds where the spin-orbit interaction is nonzero.
On the other hand, $\chi^{zz}$ is nearly unaffected by the Ising spin-orbit interaction over the whole temperature range below
the transition temperature.

If the pairing symmetry is $s$-wave in Ta$_{1.6}$Te, 
there would be a strong Rashba spin-orbit interaction so that the Knight shift is only weakly temperature dependent 
as pointed out in the previous experimental works~\cite{Terashima2024,Matsudaira2025}.
In the presence of a strong Rashba interaction, the paramagnetic depairing effect under a magnetic field is weakened and 
correspondingly the upper critical field can be large.
However, the superconductivity can be suppressed also by the orbital depairing effect, and the origin of 
the large upper critical field $H_{c2}\sim 4$T for the superconductivity in Ta$_{1.6}$Te with $T_c\simeq 1$K is to be
examined theoretically in more detail.
Besides, 
it is known that a phase modulating helical superconducting state can be stable in a Rashba superconductor under 
an in-plane magnetic field~\cite{Tada2010,Kaur2005}.
The helical state may be connected to the Fulde-Ferrell-Larkin-Ovchinnikov-like state at high magnetic field regions found in the attactive
Hubbard model on a quasicrystal~\cite{Sakai2019}. 
These issues are left for future studies.

\section*{Acknowledgements}
We thank Yuto Hirose, Kenji Ishida, Yusuke Seki, and Nayuta Takemori for fruitful discussions.
This work is supported by JSPS KAKENHI Grant No.
22K03513

\appendix
\section{System with both Rashba and Ising spin-orbit interactions}
\label{sec:rashbaising}

\begin{figure}[htb]
\includegraphics[width=6.0cm]{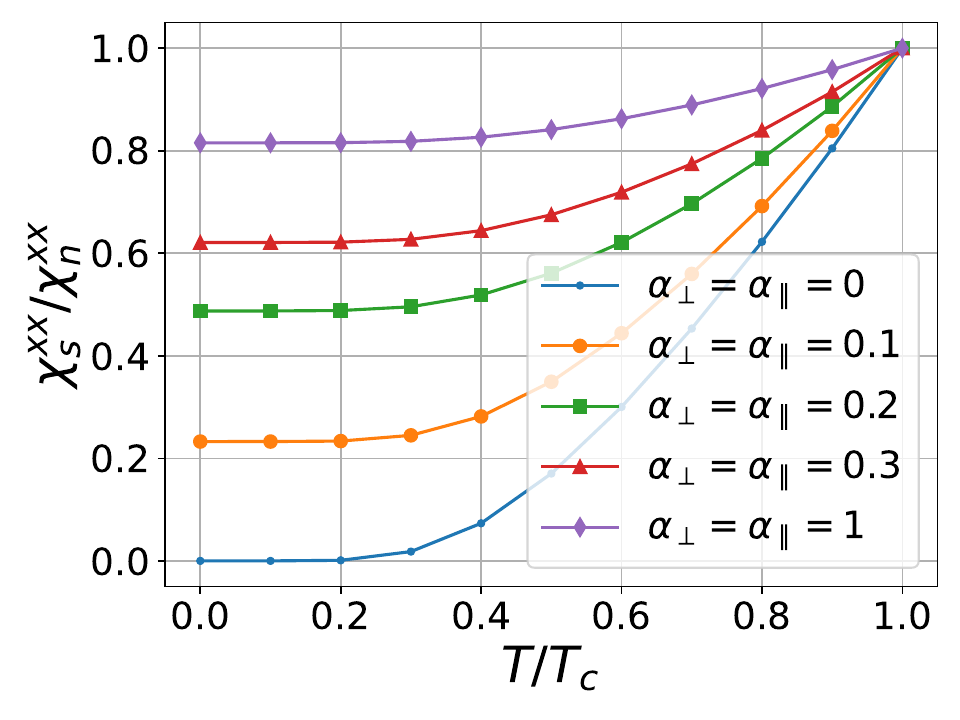}
\caption{Spin susceptibility $\chi^{xx}$ for several values of the Rashba and Ising spin-orbit couplings $\alpha_{\perp}=\alpha_{\parallel}$. 
The transition temperature is $T_c=0.1$.
}
\label{fig:chix_aRlam}
\end{figure}

\begin{figure}[htb]
\includegraphics[width=6.0cm]{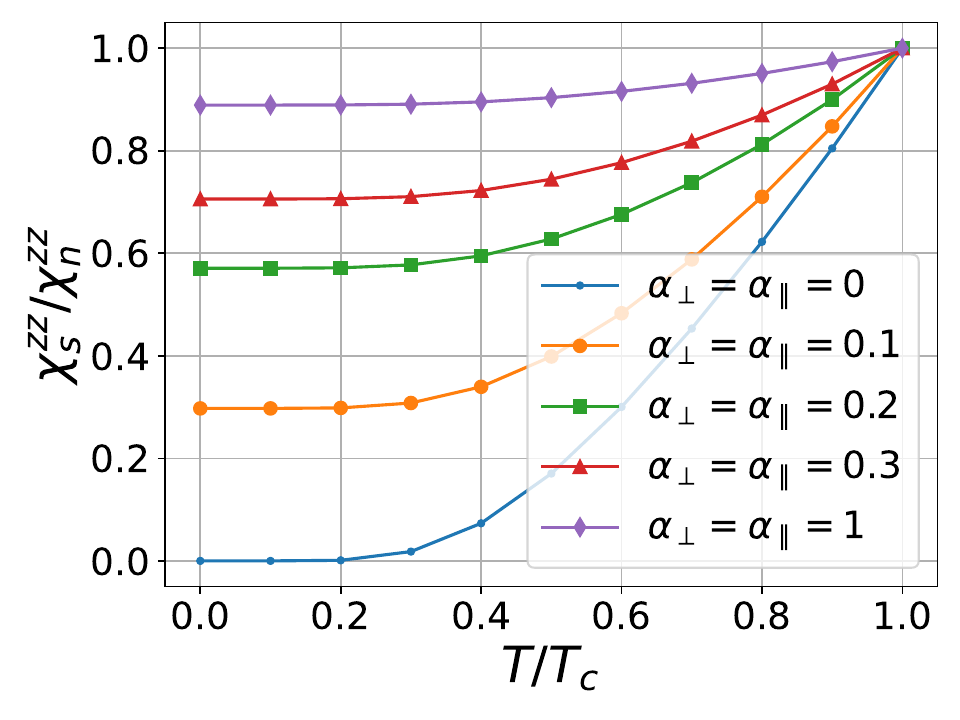}
\caption{Spin susceptibility $\chi^{zz}$ for several values of the Rashba and Ising spin-orbit couplings $\alpha_{\perp}=\alpha_{\parallel}$. 
The transition temperature is $T_c=0.1$.
}
\label{fig:chiz_aRlam}
\end{figure}

We briefly discuss the system with both the Rashba and Ising spin-orbit interactions.
As shown in Figs.~\ref{fig:chix_aRlam} and \ref{fig:chiz_aRlam},
the spin susceptibility is enhanced for both field directions.
Roughly speaking, effects of the two spin-orbit interactions simply add up.
In Ta$_{1.6}$Te, it is considered that the Rashba and Ising spin-orbit interactions coexist because its approximants do not have
the mirror symmetry along the $z$-direction and there are many bonds with no mirror symmetry within the $xy$-plane.
Therefore, we expect that 
the superconductivity is robust not only for $z$-axis magnetic fields but also for
$xy$-plane magnetic fields in a single crystal Ta$_{1.6}$Te.
Especially, the upper critical field could be huge for an in-plane magnetic field, since the orbital depairing effect would be suppressed
due to the van der Waals layered structure of Ta$_{1.6}$Te.


\bibliography{ref}



\end{document}